\documentclass[jpccfk,manuscript=article]{achemso}

\usepackage[version=3]{mhchem} 
\usepackage[utf8]{inputenc}
\usepackage[T1]{fontenc}
\usepackage{subcaption} 
\usepackage{hyperref} 

\usepackage{natbib} 

\author{Arthur Marronnier}
\affiliation{LPICM, CNRS, Ecole Polytechnique, Université Paris-Saclay, 91128 Palaiseau, France}

\email{arthur.marronnier@polytechnique.edu}

\author{Guido Roma}
\affiliation{DEN - Service de Recherches de Métallurgie Physique, CEA, Université Paris-Saclay, 91191 Gif sur Yvette, France}

\author{Marcelo Carignano}
\affiliation{Qatar Environment and Energy Research Institute, Hamad Bin Khalifa University, P.O. Box 5825, Doha, Qatar}

\author{Yvan Bonnassieux}
\affiliation{LPICM, CNRS, Ecole Polytechnique, Université Paris-Saclay, 91128 Palaiseau, France}

\author{Claudine Katan}
\affiliation{Univ Rennes, ENSCR, INSA Rennes, CNRS, ISCR (Institut des Sciences Chimiques de Rennes) – UMR 6226, F-35000 Rennes, France}

\author{Jacky Even}
\affiliation{Univ Rennes, INSA Rennes, CNRS, Institut FOTON — UMR 6082, F-35000 Rennes, France}

\author{Edoardo Mosconi}
\affiliation{Computational Laboratory for Hybrid/Organic Photovoltaics (CLHYO), CNR-ISTM, Via Elce di Sotto 8, I-06123 Perugia, Italy}

\author{Filippo De Angelis}
\affiliation{Computational Laboratory for Hybrid/Organic Photovoltaics (CLHYO), CNR-ISTM, Via Elce di Sotto 8, I-06123 Perugia, Italy}
\altaffiliation{D3-CompuNet, Istituto Italiano di Tecnologia, Via Morego 30, 16163 Genova, Italy}

\title[\texttt{achemso} demonstration]
{Influence of Disorder and Anharmonic Fluctuations on the Dynamical Rashba Effect in Purely Inorganic Lead-Halide Perovskites}

\begin{document}



\begin{tocentry}


\includegraphics[scale=0.18]{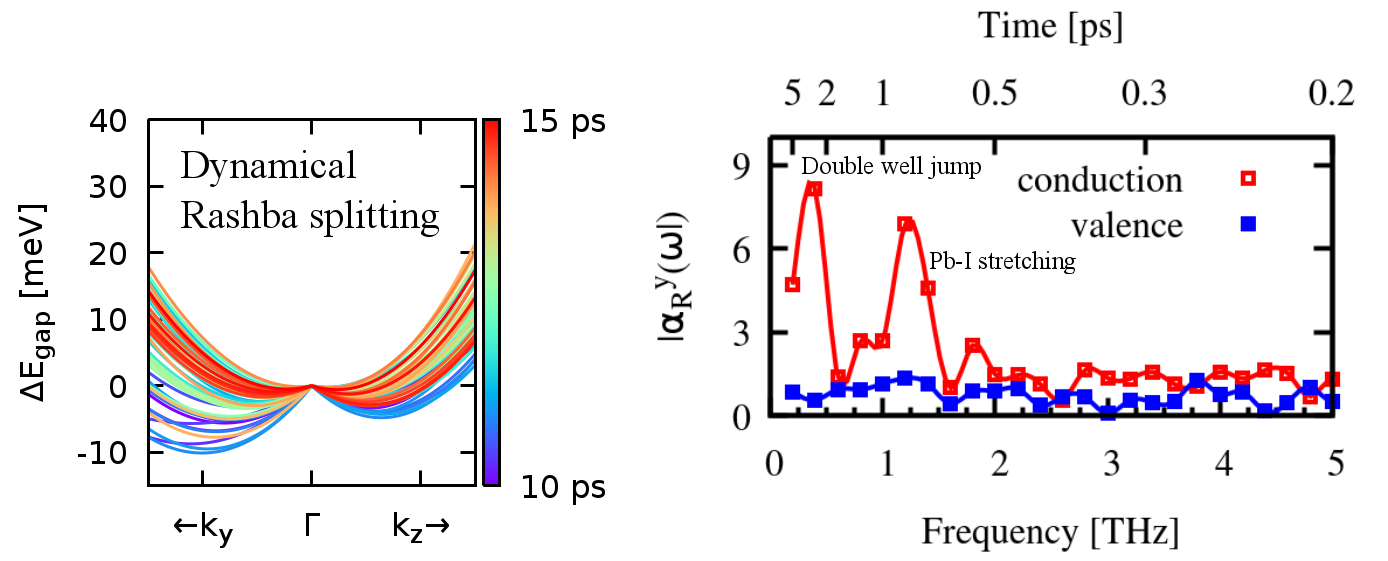}

\end{tocentry}

\newpage
\begin{abstract}

Doping organic metal-halide perovskites with cesium could be the best solution
to stabilize highly-efficient perovskite solar cells. The understanding of the
respective roles of the organic molecule, on one hand, and the inorganic
lattice, on the other, is thus crucial in order to be able to optimize the
physical properties of the mixed-cation structures. In particular, the study
of the recombination mechanisms is thought to be one of the key challenges
towards full comprehension of their working principles. Using molecular
dynamics and frozen phonons, we evidence sub-picosecond anharmonic
fluctuations in the fully inorganic CsPbI$_3$ perovskite. We reveal the effect
of these fluctuations, combined with spin-orbit coupling, on the electronic
band structure, evidencing a dynamical Rashba effect. Our study shows that
under certain conditions space disorder can quench the Rashba effect. As for
time disorder, we evidence a dynamical Rashba effect which is similar to what
was found for MAPbI$_3$ and which is still sizable despite temperature
disorder, the large investigated supercell, and the absence of the organic
cations' motion. We show that the spin texture associated to the Rashba
splitting cannot be deemed responsible for a consistent reduction of
recombination rates, although the spin mismatch between valence and conduction
band increases with the ferroelectric distortion causing the Rashba splitting.

\end{abstract}

\begin{center}
\textbf{Keywords}
\end{center}

inorganic perovskite solar cells, anharmonicity, cesium, phonons, DFT, molecular dynamics, Rashba



\newpage
\section{Introduction}

Fully inorganic metal-halide perovskites have attracted more and more
attention in the past two years as they have showed promising efficiencies
(record efficiency of 13.4\% for quantum dot devices\cite{Sanehiraeaao4204}
and of 15.07\% recently reported for a thin film\cite{Wang_CsPbI3PCE15_NatCommun2018}) and as cesium doping has proven to be a good way to improve the environmental stability of hybrid metal-halide perovskites \cite{saliba2016cesium}. Moreover, a better understanding of the physical properties of fully inorganic halide perovskites is needed in order to further understand, by contrast, the role of the organic cation in their hybrid cousins. 

In general, the enthusiasm for metal-halide perovskites can be explained by
their exceptional optoelectronic properties, whether it be their optical
properties\cite{chiarella2008combined,ogomi2014ch3nh3sn,Eperon2014,Stoumpos},
the long lifetimes of both electrons and
holes\cite{wehrenfennig2014charge,zheng2015rashba,ponseca2014organometal} and
the high mobility in these
materials\cite{cahen2014elucidating,ponseca2014organometal}. Another
remarkable feature of these materials is the fact that they present good
absorption and charge generation
properties\cite{xing2013long,stranks2013electron}, but to fully exploit this
for solar cells 
one should be able to control all the factors that limit
overall recombination rates. If the former ---absorption--- could be explained in particular by
the materials' direct band gap, for the latter ---recombination--- one expects high values for both the radiative (direct band gap) and non-radiative recombination (high density of defects). As for defects, one should note that charge separation could be actually eased in these materials through halide ionic migration\cite{lee2017direct,yang2015significance} which could either favour exciton screening\cite{even2014analysis} or give birth to local screening domains\cite{ma2014nanoscale,Quarti2016struct}.

Concerning radiative recombination, which has been indeed shown to be
relatively high\cite{DaviesNatCommun2018}, one should take into account the
interplay of spin and orbital degrees of freedom, which are of important
magnitude in these materials because of the presence of the heavy lead
atoms. In particular, the giant spin-orbit coupling (SOC) that was reported in
these materials \cite{even2013importance} is expected to be at the origin of
Rashba-like splittings\cite{Kim6900,Even_Rashba,Amat,Brivio_PRB14}. Such
splittings correspond to the lift of the electronic bands' spin degeneracy in
the presence of SOC and time reversal symmetry when the inversion symmetry is
broken in the crystal\cite{zhang2014hidden,Kepenekian}. Assuming long-range
polar distortions of the perovskite lattice, it has been speculated that these
band splittings can drastically impact the recombination rates by limiting
direct transitions between the valence and conduction bands. This impact was
theoretically estimated by Zheng \textit{et al.} to contribute to a reduction
of the recombination rates reaching up to two orders of magnitude
\cite{zheng2015rashba}. However, the existence of long-range polar distortions
of iodide-based perovskite lattices is still debated,  the influence of
Rashba-like spinor band splittings could be more subtle and rather related to
local lattice distortions. Moreover, very recent
works\cite{ZhangJPCLett2018,ZhangACSELett2018} question the role of Rashba
splitting in MAPbI$_3$ (MA=methylammonium), where the inversion symmetry breaking is associated to the orientation of methylammonium ions. The Rashba effect can influence carrier transport in halide perovskite also by modifying the carrier mobility, due to modified electron-phonon scattering.\cite{YuJPCLett2016,Kang_CsPbX3carriermobility_PRApp2018}

Etienne \textit{et al.} \cite{Etienne} investigated by DFT-based molecular
dynamics the interplay of electronic and nuclear degrees of freedom in the
prototype MAPbI$_3$ perovskite and revealed a dynamical Rashba effect. They
reported the influence of temperature and found a spatially local Rashba
effect with fluctuations at the subpicosecond time scale, that is to say on
the scale of the MA cation motion. Although this time scale is much smaller
than the carrier lifetimes, apart from directly affecting radiative
recombination probabilities, it could influence the kinetic path for
electron-hole recombination.\cite{Hutter_indirectRec_ACSELett2018}
It is worth pointing out that this
numerical demonstration by Etienne and coworkers was based on MAPbI$_3$ structures preserving centrosymmetry
on the average. They noticed that the Rashba splitting can be quenched when
reaching room temperature but also when using larger supercells (up to 32 MAPbI$_3$
units, i.e., 3 nanometers cells) representing a higher and more realistic
spatial disorder.  
The effect is still debated,\cite{Frohna_noRashbaMAPI_NatCommun2018}
but an experimental evidence of dynamical Rashba splitting in MAPbI$_3$ was
recently reported.\cite{Niesner_expDynRashba_PNAS2018}

The local and dynamical nature of polar distortions may weaken the influence
of Rashba-like spinor splittings by comparison to long range and static polar
distortions. However the lack of long range correlations between local polar
distortions could be compensated by the unusually strong amplitudes of the
atomic motions. The strong anharmonicity of the perovskite lattice is a
general feature of this new class of semiconductors\cite{Katan_riddles}, that
was pointed out experimentally very early by inelastic neutron scattering in
the context of inorganic halide  perovskites\cite{Neutron_Cl} and that shall
give rise to at least two characteristic experimental signatures: large and
anisotropic Debye-Waller factors in diffraction
studies\cite{Trots,hutton1979high} and a so-called quasielastic central peak
observed either in inelastic neutron  or Raman scattering studies
simultaneously with highly damped
phonons\cite{Even_disorder,yaffe2017local}. However, the strong perovskite
lattice anharmonicity is not expected to affect only zone center polar optical
modes, but also acoustic modes or optical modes located at the edges or the
Brillouin zone and related to non-polar
antiferrodistortions\cite{Katan_riddles}. The previously mentioned
experimental signatures (Debye-waller factors, phonon damping and central
peaks by inelastic neutron or Raman scattering studies) can hardly be
considered as unambiguous experimental proofs of the existence of strongly
anharmonic polar fluctuations. Nowadays direct experimental investigations of
the dielectric response give useful indications about the influence of lattice
polar distortions\cite{Marronnier2018179,anusca2017dielectric} and the
importance of the Fröhlich interaction\cite{Sender_Mater} for electron-phonon
coupling processes\cite{fu2018unraveling}, but do not directly probe the
anaharmonicity of polar distortions. Numerical simulations are therefore still
useful tools that already allowed showing the presence of anharmonicity
features in CsPbI$_3$\cite{Marronnier,Marronnier_ACSNano}, leading to symmetry
breaking minimum structures in the high temperature phases both at the edges
and at the center of the Brillouin zone. 
MD simulations for CsPbBr$_3$ also suggested that the fluctuations in this material are mostly due to head-to head Cs motion and Br face expansion happening on a few hundred femtosecond time scale\cite{yaffe2017local}.

In that sense, large polar fluctuations of the perovskite lattice at the local scale may lead to two main effects: Rashba-like spinor splittings and strongly anharmonic polarons related to the Fröhlich interaction. We focus in the present contribution on the former aspect.


\section{Results and Discussion}
\label{results}

In this article, we aim to analyze the Rashba effect induced by the anharmonic
double well and the influence of symmetry breaking that has been evidenced in
the cubic phase of CsPbI$_3$.\cite{Marronnier,Marronnier_ACSNano} In fact it
was shown, using the frozen phonon method, that the highly symmetric cubic phase can be distorted to form two lower-symmetry structures with a slightly lower total energy (by a few meV). These two distorted structures, that we will call in the rest of the article $A^+$ and $A^-$, have no inversion symmetry and correspond to the two minimum structures of the double well-instability discussed in Ref.\citenum{Marronnier}.
They correspond to opposite ferroelectric distortions ($\eta > 0$ and $\eta <
0$) along a soft polar eigenmode of the centrosymmetric $Pm\overline{3}m$
("S") structure, represented in Figure \ref{Rashba_spatial}a, along the $x$-direction. 
Note that with rotational symmetry similar studies could be done on the two other eigenmodes corresponding to distortions along the two remaining Cartesian axes (y and z given our labeling).
The first aim of the study here is to look at the possible formation of "$A^+$ domains" and "$A^-$ domains", both in space (supercells) and in time (Car-Parrinello molecular dynamics "CPMD").

Then, we analyze in detail the dynamical Rashba effect induced by the time dynamics of the oscillations between structure $A^+$ and structure $A^-$ through the highly symmetric structure S ($\eta = 0$). This study is done on CPMD trajectories obtained from Ref. \citenum{carignano2017critical}

\subsection{Spatial disorder}
\label{test}

\begin{figure}[!h]
\begin{center}
  \includegraphics[scale=0.2]{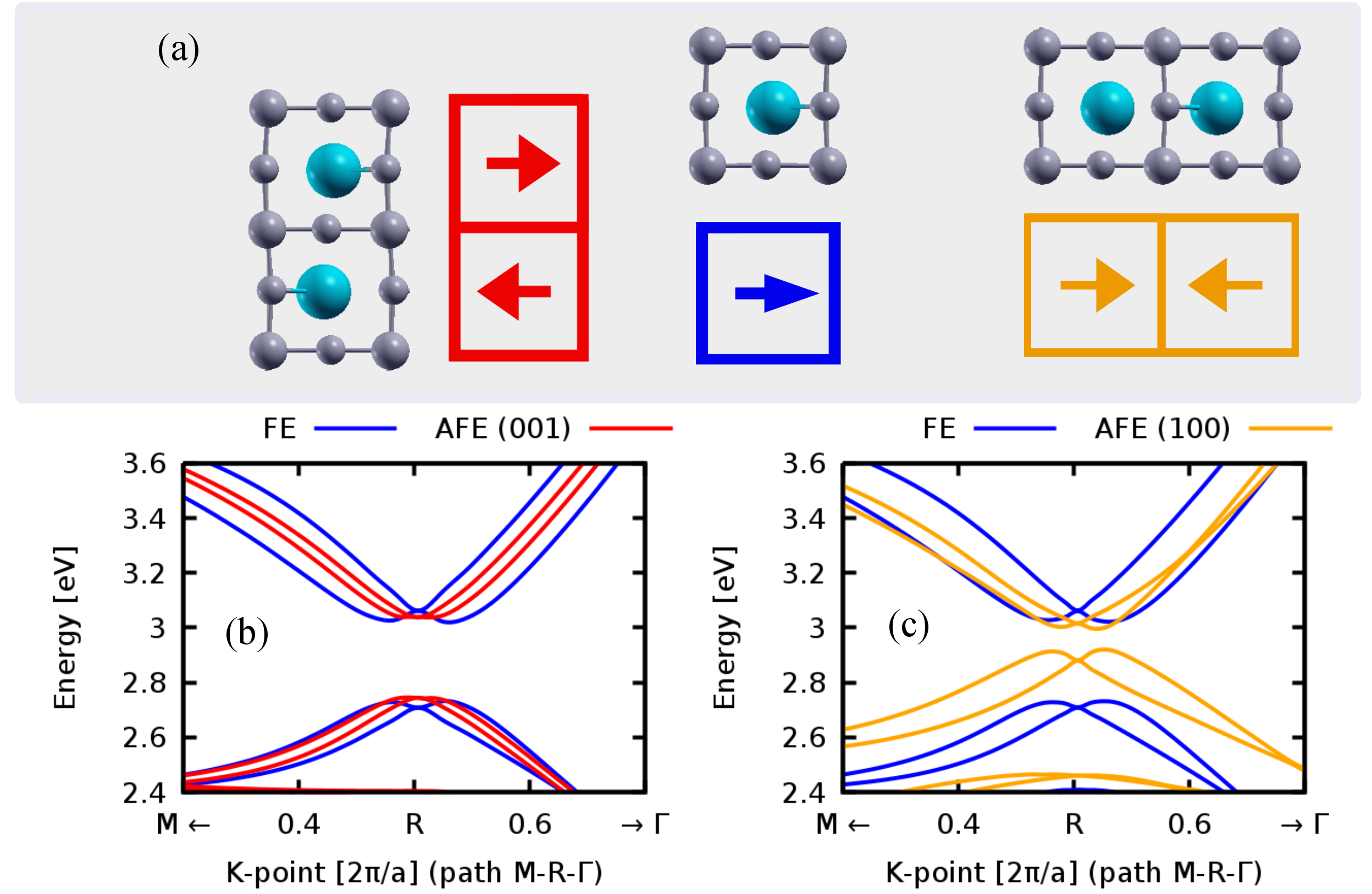}
  \caption{(a) Sketch of the atomic structure and schematic representations of the corresponding unit cells used to study the
    influence of spatial domains $A^+$/$A^-$ or "x up"/"x down" on the
    electronic band structure (arrows represent displacement of Cs, the blue atom). (b,c) Electronic band structure (including SOC) of the 2$\times$1$\times$1 (orange) and 1$\times$1$\times$2 (red) anti-ferroelectric (AFE) configurations, both compared to the ferroelectric (FE) one (blue). The labelling of the k-points is referred to the BZ of the cubic unit cell.}
  \label{Rashba_spatial}
  \end{center}
\end{figure}

First, the aim is to investigate the influence of spatial $A^+$/$A^-$ domains
in CsPbI$_3$ on its electronic band structure, in particular in terms of Rashba effect. Given that the eigenvector under study mostly corresponds to a distortion along one axis 
(here we considered the x-axis) we built supercells by doubling the unit cell
either in the x direction or in the z direction. These 2$\times$1$\times$1 and
1$\times$1$\times$2 supercells are built putting together 2 unit cells: 1 unit
cell in configuration $A^+$ (or x up), and another one in configuration $A^-$ (or x down). These supercells, representing modulated structures with the smallest possible period, are schematically shown in Figure \ref{Rashba_spatial}a.

The Rashba splitting obtained at the R point for a unit cell, in the
minimum, symmetry-breaking structure is shown in Figure
\ref{Rashba_spatial}, the blue curve in panels a and b. When doubling the cell along z (resp. x), the R point
folds onto the S point (resp. T point), using the orthorhombic convention (in
the figure we kept the labels of the cubic unit cell) . For this
ordered, static reference structure we find energy splittings of 42 meV (57 meV with
LDA) and 24 meV (40 meV with LDA) between the $\Gamma$ point and, respectively, the
conduction band minimum and valence band maximum. In order to give an estimate of the Rashba splitting taking into account the effect both in energy and in the k-space, we calculated the commonly used $\alpha$ parameter as defined in Ref. \citenum{Etienne}:

\begin{equation}
    \alpha_{C,V} = \frac{\Delta E_{C,V}}{2\Delta k_{C,V}}
\end{equation}

where $\Delta E_{C,V}$ is the energy difference between the first (resp. last)
two bands of the conduction (resp. valence) bands and $\Delta k_{C,V}$ the
splitting of the minimum (resp. maximum) in the k-space. For the reference
static, highly ordered structure we found $\alpha$ values of 3.04 (4.09 with
LDA) eV.$\mathring{A}$ and 2.75 (2.01 with LDA) eV.$\mathring{A}$ for
the conduction and the valence bands, respectively. These values are comparable
to those reported in Ref. \citenum{Etienne} in the case of polar MAPbI$_3$: 3.17 eV.$\mathring{A}$ and 1.17 eV.$\mathring{A}$ respectively. Note that so far the highest values found in ferroelectric materials for the Rashba parameter are 4.2-4.8 eV.$\mathring{A}$ for GeTe \cite{alpha_max1,alpha_max2}.
\begin{figure}[!h]
\begin{center}
   \includegraphics[scale=0.3]{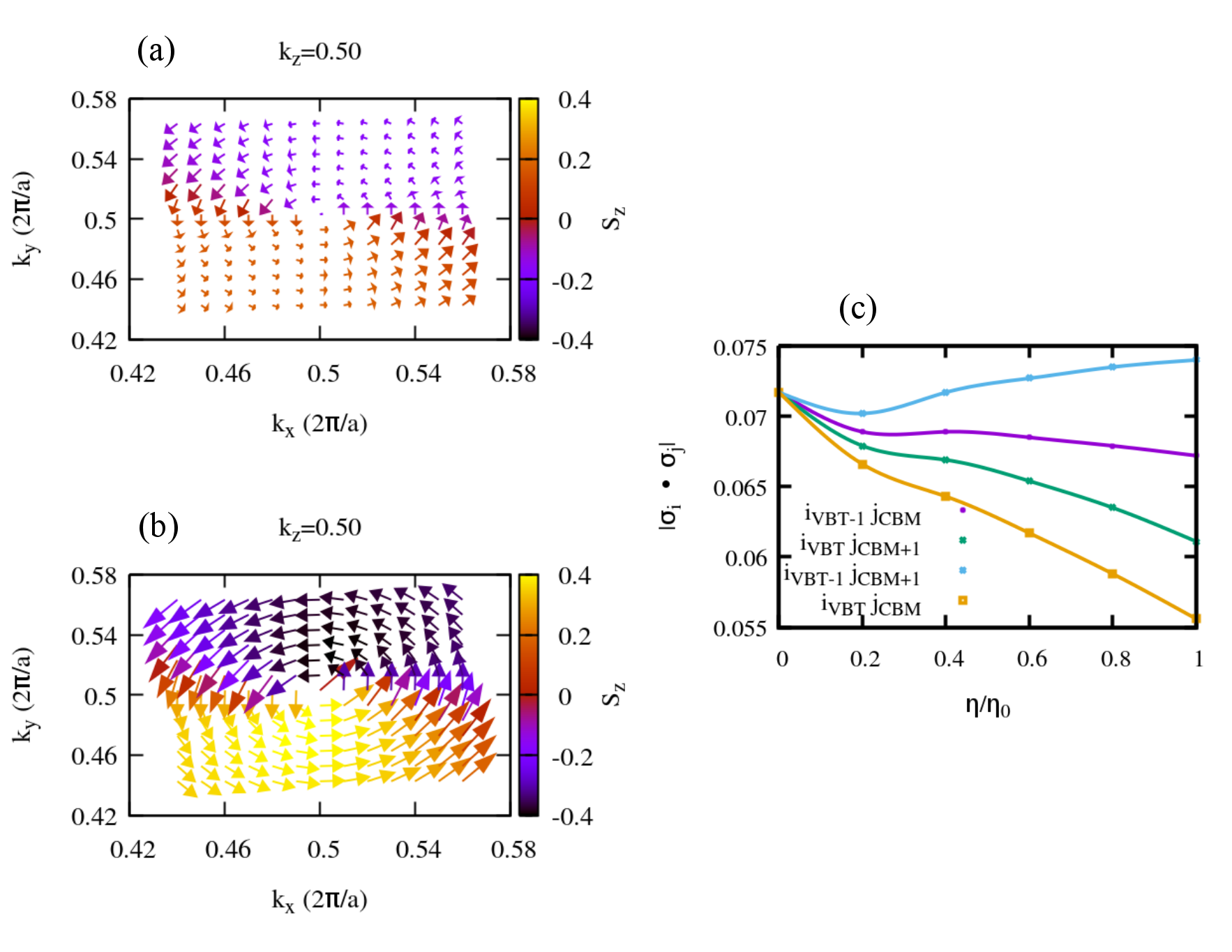}
  \caption{Spin textures for the last occupied valence band (a) and the first
    empty conduction one (b) in a cube with side 0.06$\frac{\pi}{a}$
    surrounding the R point of the BZ. A section with z=0.5 is shown. The
    third panel (c) shows the integral of four scalar products between the spin
    vector expectation values of a valence and a conduction state as a
    function of the polar distortion leading from the centrosymmetric
    $Pm\overline{3}m$ "S" structure ($\eta/\eta_0=0$) to the distorted minimum energy
    "A+" structure ($\eta/\eta_0=1$), where the inversion symmetry is broken. The
    last two occupied states (VBT-1 and VBT) and the first two empty ones (CBM
  and CBM+1) are considered.} 
  \label{SpinTexture}
  \end{center}
\end{figure}


The results for the two modulated structures are shown in Figures \ref{Rashba_spatial}b and \ref{Rashba_spatial}c. Whereas no Rashba effect is found in the case of a modulation orthogonal to the direction of symmetry breaking, a band splitting around the valence band maximum and the conduction band minimum is found for a modulation parallel to the direction of symmetry breaking. We found $\alpha$ values of 1.98 (0.58 with LDA) eV.$\mathring{A}$ and 2.97 (2.18 with LDA) eV.$\mathring{A}$ for the conduction and valence bands, respectively. 

In general, the Rashba splitting in the band structure of a two-dimensional system results from the combined effect of atomic spin-orbit coupling and asymmetry of the potential in the direction (here x) perpendicular to the two-dimensional plane, causing a loss of inversion symmetry. 
In the case of a modulation orthogonal to the direction of symmetry breaking
(1$\times$1$\times$2 supercell), the symmetry along x is respected on average:
the inversion symmetry is kept and the Rashba splitting vanishes. We expect then that the quenching of the Rashba effect results from a competition between parallel and orthogonal modulations: the former keeps the Rashba effect, while the latter tends to cancel it.

According to Rashba model hamiltonians the spin texture is such that the top of the
valence band and bottom of the conduction band of the splitted bands have
opposite spin orientations, reducing thus the recombination probability at the
band gap. It has been recently shown, through full ab initio calculation of
the spin texture~\cite{ZhangJPCLett2018}, that this picture is not realistic
for MAPbI$_3$, because the spin mismatch between the highest valence band and
the lowest conduction bands is small and, as such, not expected to reduce the
recombination rate in a significant way. In MAPbI$_3$ inversion symmetry is
broken by a combination of orientation of the methylammonium molecular ion and
the distortion it induces in the inorganic network. It is then difficult to
disentangle the two effects. CsPbI$_3$ is a useful playground to single out
the effect of the polar distortion. Calculating the spin texture in a region
of the BZ close to the R point of the cubic structure we found that, as for
MAPbI$_3$, the spin orientations are similar around the R point for the last
occupied and the first empty states. We show two maps in panels (a) and (b) of Figure \ref{SpinTexture}.  
In order to quantitatively assess the collinearity of the spins of valence and
conduction states close to the band gap we show, in panel (c) of Figure
\ref{SpinTexture}, the integral of the modulus of the scalar product of valence/conduction spin vectors in a cube of size 0.06$\frac{\Pi}{a}$ around the R-point; the four possible combinations between the last two valence bands and the first two conduction ones are presented. The curves show that all the four products have similar order of magnitudes, that the lowest value is the one between the last occupied band and the first empty one and, finally, that the polar distortion does indeed enhance the spin mismatch of three out of four valence/conduction pairs.

\subsection{Dynamical structural fluctuations}
\label{time_subsec}

Next, we analyze in detail CPMD trajectories of cubic CsPbI$_3$ in the light of our findings on the double well potential energy surface. The trajectories were computed by Carignano \textit{et al.} in the framework of a study \cite{carignano2017critical} of the anharmonic motion of the iodine atoms in CsPbI$_3$ and MAPbI$_3$, where they showed that, at variance with FAPbI$_3$, these two perovskite structures are expected to have a deviation from the perfect cubic unit cell at any time of the MD, with a probability very close to 1. This hints towards the interpretation that the $Pm\overline{3}m$ symmetry can be seen as a time average, including for CsPbI$_3$. This phenomenon had already been reported for MAPbI$_3$ in earlier studies \cite{Quarti2016struct}, where it was evidenced that the system strongly deviates from the perfectly cubic structure in the sub-picosecond time scale.

The molecular dynamics simulation were performed at 370 K under NPT-F conditions, which allow volume fluctuations by
changing the supercell edges and angles. The temperature was controlled by a
Nose-Hoover thermostat with three chains, and the pressure was controlled by
the Martyna’s barostat \cite{martyna1996explicit}. The time constant for both,
the thermostat and barostat, was set at 50 fs. The system used for
CsPbI$_3$ has 320 atoms (4$\times$4$\times$4 supercells).


\begin{figure}[!h]
\begin{center}
  \includegraphics[scale=0.35]{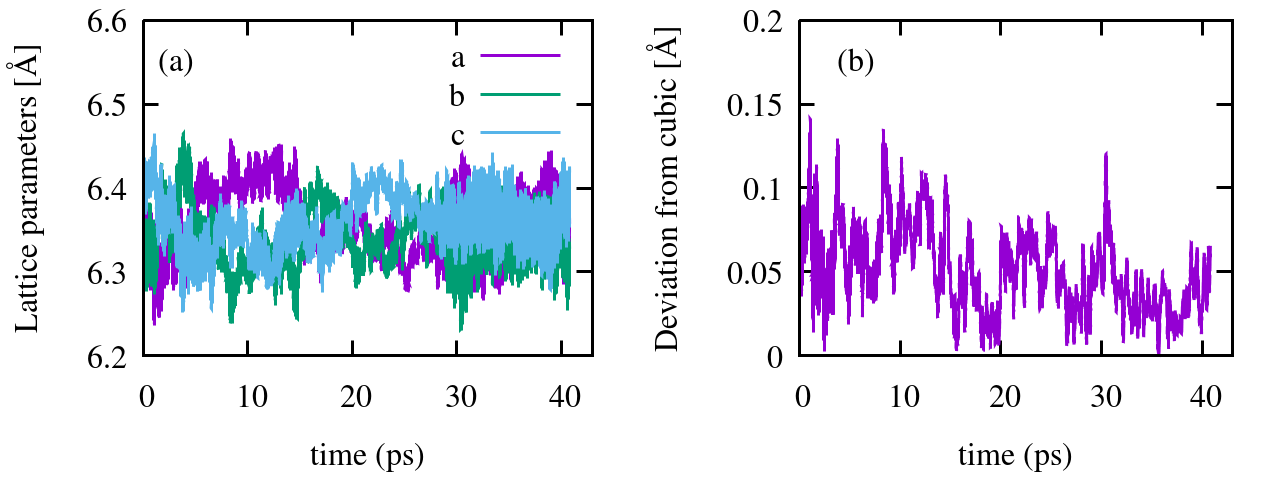}
  \caption{(a) Lattice parameters fluctuations along the CPMD trajectory at 370 K. (b) Fluctuations of the deviation from the average pseudocubic structure, as defined in Eq. \ref{eq_distance_pseudo}.}
  \label{panel_2}
  \end{center}
\end{figure}

In Figures \ref{panel_2}a we show the lattice parameters fluctuations versus
time. In particular, from this first simple analysis we can infer that the
structure fluctuates around a cubic structure: the difference between the
lattice parameters stays below 3\%. Even though the structure is not perfectly
cubic on average (see Table \ref{avg_latt_MD}), the deviation from the average
pseudocubic lattice structure (Figure \ref{panel_2}b) is smaller than
1\%. This deviation $d$ is a cartesian distance obtained as:

\begin{equation}
\label{eq_distance_pseudo}
d (t) = (\sum_{i=1}^{3} (x_i(t)-\bar{x_i})^2)^{\frac{1}{2}}
\end{equation}

where $x_i$ are the 3 lattice parameters and $\bar{x_i}$ their time average over the whole trajectory.

\begin{table}
\begin{center}
  \begin{tabular}{c | c | c |}

    In Angstroms & 370 K &  450 K\\ \hline
    a & 6.358 & 6.372   \\
    b & 6.338 & 6.391   \\
    c & 6.358  & 6.361     \\
  \end{tabular}
\caption{Average lattice parameters (in Angstroms) along the CPMD trajectories at 370 K and 450 K.}
  \label{avg_latt_MD}
\end{center}
\end{table}


In order to analyze the MD trajectories in the light of the aforementioned double well instability, we project these trajectories onto two kind of structures: the perfectly cubic symmetric structure ("S") and the symmetry breaking structures $A^+$ and $A^-$. The chosen approach is to study the radial distribution function of the cesium-lead pairs during the MD simulation and to compare it to our two reference structures.

\begin{figure}[!h]
\begin{center}
  \includegraphics[scale=0.75]{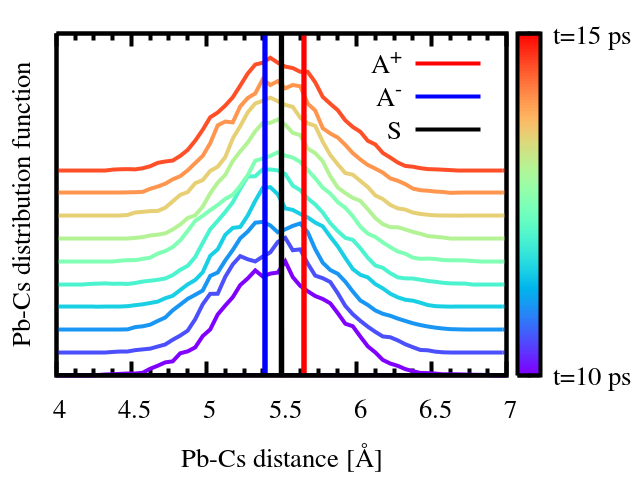}
  \caption{Distribution function of the cesium-lead pairs' distances along the MD trajectory. Here the references (vertical lines) correspond to the distances for structures S, $A^+$ and $A^-$, weighted by the ratio between the lattice parameters.}
  \label{PbCs_10to15ps_mini_vol}
  \end{center}
\end{figure}

Figure \ref{PbCs_10to15ps_mini_vol} focuses on averages over 0.5 ps intervals. At this time scale, our double well references seem to explain very well how the system explores the energy landscape. Whereas some intervals show a distance peak corresponding to the distance in the average pseudocubic structure S, for instance the [11-11.5 ps] interval shows two peaks centered on both minimum structures $A^+$ and $A^-$. This means that within 0.5 ps the structure has enough time to explore the whole double well. We think that this is the most appropriate time-scale to evidence the double well instabilities.

\subsection{Dynamical Rashba effect}
\label{Rashba_subsec}

We now focus on the dynamical Rashba effect possibly ensuing from the nuclear dynamics exposed above. We expect to find in CsPbI$_3$ an effect similar to what was evidenced for MAPbI$_3$ for which the spatially local Rashba splitting was found to fluctuate on the subpicosecond time scale typical of the methylammonium cation dynamics \cite{Etienne}.

To investigate this effect, we calculate the electronic band structure, including spin-orbit coupling, at different snapshots along the trajectory. Given the results of the Pb-Cs distance analysis, we chose to focus these calculations on the [10-15 ps] interval in which we chose 50 regularly distributed snapshots (hence separated by 100 fs from each other) in order to better capture the sub pico-second dynamics. For each snapshot, we used the MD structure of the 4$\times$4$\times$4 supercells (we remind the reader that the cell's atomic positions, lattice parameters and angles vary) and derived its electronic band structure (see the Methods section). These calculations for 4$\times$4$\times$4 supercells follow the guidelines of those previously done for MAPbI$_3$ \cite{Quarti2016struct}.

\begin{figure}[!h]
\begin{center}
  \includegraphics[scale=0.18]{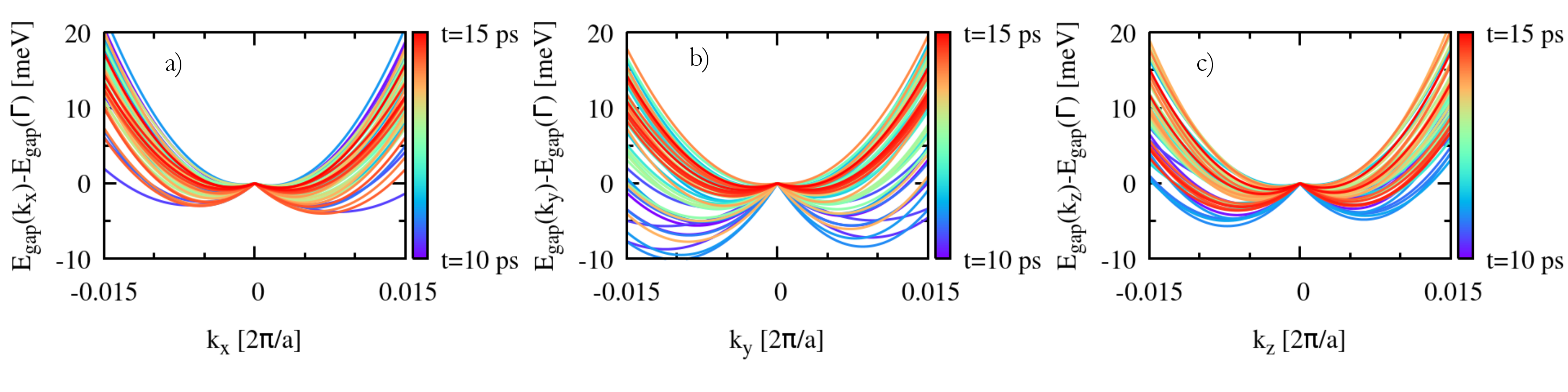}
  \caption{Differences between the gap at finite {\bf k} and at $\Gamma$ for
    the 50 snapshots chosen along the MD trajectory. The three panels a-c
    refer to the three directions, {\bf k}$_x$,  {\bf k}$_x$ {\bf k}$_z$. This difference is 0 at $\Gamma$ by construction (see Eq. \ref{eq_gap}). }
  \label{Eg_50}
  \end{center}
\end{figure}

The electronic band structure calculations are done at 7 {\bf k} points of the
Brillouin zone and, from these, parabolic bands in the vicinity of the
$\Gamma$ point were obtained. (For further details see the Supporting Information).
In Figure \ref{Eg_50} we plot, for each snapshot $i$ of the 50 structures chosen
in the MD trajectory and for each {\bf k} point, the normalized energy
difference defined as:

\begin{equation}
    \Delta E_{gap}^i(k) = [CBM^i(k) - VBT^i(k)] - [(CBM^i(\Gamma) - VBT^i(\Gamma))]
    \label{eq_gap}
\end{equation}

where CBM is the conduction band minimum and VBT the valence band top. This is
necessary as the cell is variable along the trajectory: the fluctuations on
the gap value, which are large with respect to the Rashba splitting, would
mask it otherwise. The corresponding plots for the valence and conduction
bands as well as the evolution of the bandgap over time are presented in the
supporting information.

These results show that 100 fs is a good estimate of the timescale of the Rashba effect dynamics. Moreover, on average we see a band gap shift to the Y direction, the band gap being reduced by 1.3 meV on average. Taking the extreme case, we can infer that the amplitude of the oscillations in the 5 ps timescale is around 10 meV. 
Further analysis of the time coherence of the Rashba effect are provided in the supporting information file through time correlation functions. 

Figure S1 shows that this is mostly due to a Rashba splitting happening at the
CBM rather than at the VBT. This is coherent with the fact that the most
relativistic atom, Pb, is mostly contributing to the conduction band and with
what was previously reported for MAPbI$_3$ \cite{Quarti2016struct} as well.

\begin{figure}[!h]
\begin{center}
  \includegraphics[scale=0.5]{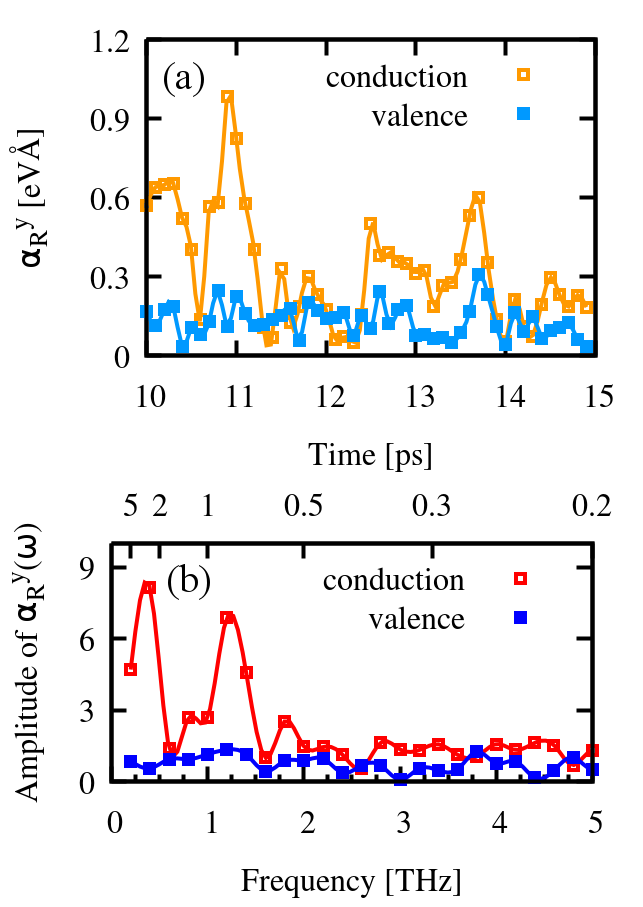}
  \caption{(a) $\alpha_R$ parameter for the conduction and valence bands (in the
    {\bf k}$_y$ direction) versus time. (b) Fourier transform of the Rashba $\alpha$
    parameter shown in panel (a).} 
  \label{alpha_fourier}
  \end{center}
\end{figure}

Figure \ref{alpha_fourier}a represent the oscillations of the Rashba effect in this interval through the previously defined $\alpha$ parameter. This result confirms that the Rashba effect is much more substantial for the conduction band than for the valence band, and oscillates with values close to 1 eV.$\mathring{A}$. Even though this is smaller than in the static case (values around 3 eV.$\mathring{A}$), this means that the effect is still sizable despite the disorder induced by temperature and the large investigated supercell.

It is interesting to further look into these oscillations through a Fourier analysis (Figure \ref{alpha_fourier}b) which reveals the existence of two main frequency components : 

\begin{itemize}
    \item A 2.5 ps component (frequency: 0.4 THz or 13 cm$^-1$) which could
      correspond to the jump through double well and thus to the slow motion
      of the Cs cation;
    \item A 0.8 ps component (frequency: 1.2 THz or 40 cm$^-1$) corrresponding to the phonon modes usually associated to the Pb-I stretching (around 20-40 cm$^{-1}$).
\end{itemize}

We further compare the order of magnitudes of these
oscillations to those obtained in a similar study lead on hybrid perovskite
MAPbI$_3$ by Etienne \textit{et al.}.\cite{Etienne}. We report in Table
\ref{table_alpha} the corresponding values for the apolar structure of
Ref. \citenum{Etienne}, because a Cs atom has no permanent dipole moment. Note
that nevertheless, for the polar structure, the largest $\alpha$ value
reported in Ref. \citenum{Etienne} (10.36) is even smaller than the largest
one reported for the non-polar structure. One needs to keep
in mind that we have here very large supercells compared to what was used in
that study. The conclusion we can draw from this comparison is that we observe
a sizable dynamical Rashba effect even with large supercells and the absence
of the organic molecule, which in general is a possible source of symmetry
breaking in these halide perovskite structures. The fact that the order of
magnitude of the dynamical Rashba splitting is similar for MAPbI$_3$ and
CsPbI$_3$ is in agreement with the recent observation of similar
recombination kinetics for both compounds.\cite{Dastidar_CseqMA_ACSEnLett2017}

\begin{table}
\begin{center}
  \begin{tabular}{c | c | c |}

    Number of formula units & $\alpha$ for hybrid MAPbI$_3$  &  $\alpha$ for inorganic CsPbI$_3$ \\
    & from Ref. \citenum{Etienne} (eV.$\mathring{A}$) & from our results (eV.$\mathring{A}$)\\ \hline
    1 & 12.48 &    \\
    4 & 3.86 &    \\
    32 & 2.19 &     \\
    64 &  & 0.96    \\
  \end{tabular}
\caption{Maximum value of the $\alpha$ oscillations.}
  \label{table_alpha}
\end{center}
\end{table}

\section{Concluding remarks}

In summary, we investigated the effect of spatial and temporal disorder 
on the Rashba splitting in the cubic $\alpha$ phase of inorganic halide
perovskite CsPbI$_3$. The analysis focused on the fluctuations of the Rashba
$\alpha$-parameter ---a measure of the Rashba band splitting--- along a
molecular dynamics trajectory for a large supercell\cite{carignano2017critical}. 
Our results highlight a dynamical Rashba effect similar to the one previously observed for hybrid organic-inorganic halide
perovskites\cite{Etienne}, which persists in spite of the
quenching effect coming from spatial disorder in this relatively large simulation cell (320 atoms). 
Some low-frequency vibrational modes of the system, and in particular the anharmonic behavior, which has been shown to originate from the double well potential energy landscape of a polar optical phonon\cite{Marronnier}, contribute to the spatial extension of the Rashba effect. This is confirmed by the Fourier analysis of the Rashba $\alpha$-parameter fluctuations.

An expected consequence of the Rashba effect is the reduction of the carriers recombination rate\cite{zheng2015rashba} and consequent enhancement of their radiative lifetime; however, this effect is submitted to a specific spin texture which has been recently shown not to occur in MAPbI$_3$. Our calculated ab initio spin textures in CsPbI$_3$ suggest that, although a reduction of the recombination rate due to Rashba splitting is indeed expected, its effect is not as large as previous model calculations had predicted.

\section{Methods}

\subsection{Density functional theory for band structure calculations}


\label{meth_subsec}
In this study we started from the minimum reference structures obtained in
Ref. \citenum{Marronnier} which were optimized with the PBE functional letting both
the lattice parameters and the atomic positions relax, keeping the cell's
angles fixed. These structures are thus slightly orthorhombic (0.6\%
distortion). The total energy gain due to the polar distortion is
  smaller with PBE (3.4 meV) than with LDA, as reported in
  Ref. \citenum{Marronnier_ACSNano} with and without SOC and slightly different settings, but with an analogous energy profile.


For these reference structures, geometry optimizations and force calculations
were performed with spin-orbit coupling (SOC). Fully relativistic
pseudopotentials were used, with the Cs $[5s^25p^66s^1]$, I $[5s^25p^5]$ and
Pb $[5d^{10}6s^{2}6p^2]$ electrons treated as valence states. The choice of 14
electrons for Pb and 9 for Cs was previously discussed\cite{Marronnier}. PBE
pseudopotentials are ultrasoft ones and  were used
with a wave function energy cutoff of 25 Ry (200 Ry for the charge density). LDA ones
were norm conserving and were used with a 70 Ry wave function cutoff energy. 

In order to investigate the effect of spatial and dynamical disorder on the
Rashba effect, we proceeded with the following two steps:
\begin{itemize}
\item as for the study of the spatial domains, the band structure calculations
  of the constructed 2$\times$1$\times$1 and 1$\times$1$\times$2 supercells
  were performed with fully relativistic pseudopotentials (for Pb and I) with
  PBE and SOC. The same calculations were made also with LDA and showed qualitatively
  similar results.

\item in order to study the dynamical Rashba effect from CPMD, the band
  structure calculations were performed with the same fully relativistic US
  pseudopotentials using the PBE xc functional, in coherence with the CPMD
  calculations from which MD trajectories were taken
  \cite{carignano2017critical}, which were done using PBE as well (with the
  CP2K code). 
\end{itemize}

Spin textures were calculated by obtaining the expectation value of the spin
operators in the three cartesian directions on a single particle Kohn-Sham
wave function ($S_i^\alpha=\langle \psi_i|\sigma_\alpha|\psi_i\rangle$,
where $\sigma_\alpha ,\alpha=1,2,3$ are the three Pauli spin matrices,
$\psi_i$ is a two component spinor eigenfunction).  
The Brillouin zone was sampled with $\Gamma$-centered Monkhorst-Pack meshes
\cite{monkhorst1976special} with subdivisions of 8$\times$8$\times$8 k-points
for unit cells, and corresponding sampling when doubling the cell in $x$ or
$z$ directions. The molecular dynamics snapshots were sampled with the
$\Gamma$ point only. 

\section{Associated Content}

The authors declare no competing financial interest.

\acknowledgement

Dr. Arthur Marronnier's PhD project was funded by the Graduate School of École des Ponts ParisTech and the French Department of Energy (MTES).
HPC resources of TGCC and CINES were used through allocation 2017090642 and x20170906724 GENCI projects.
The work at FOTON and ISCR was funded by the European Union’s Horizon 
2020 program, through a FET Open research and innovation action under the grant agreement No 687008.

\begin{suppinfo}

We provide in the supporting information additional results on the effect of
dynamical disorder on the conduction band minimum and valence band top
throughout the chosen 5 ps interval of study for our band structure
calculations. We also provide an analysis of the time correlation of the
Rashba parameter $\alpha$ (or $\alpha_R$) and of the spin textures.

\end{suppinfo}

\bibliography{sample_clean}

\end{document}




\newpage

\section{Introduction}
We provide, in this Supporting Information file, further data on the band
structure of the snapshots extracted from the MD simulation and some further
analysis of the Rashba parameter $\alpha_R$ and other relevant quantities
related to the spin texture throughout the 5 ps of the simulation on which we
focused our study.

\section{Band structure}
\label{BandStructure}
\begin{figure}[!h]
\begin{center}
\includegraphics[scale=0.25]{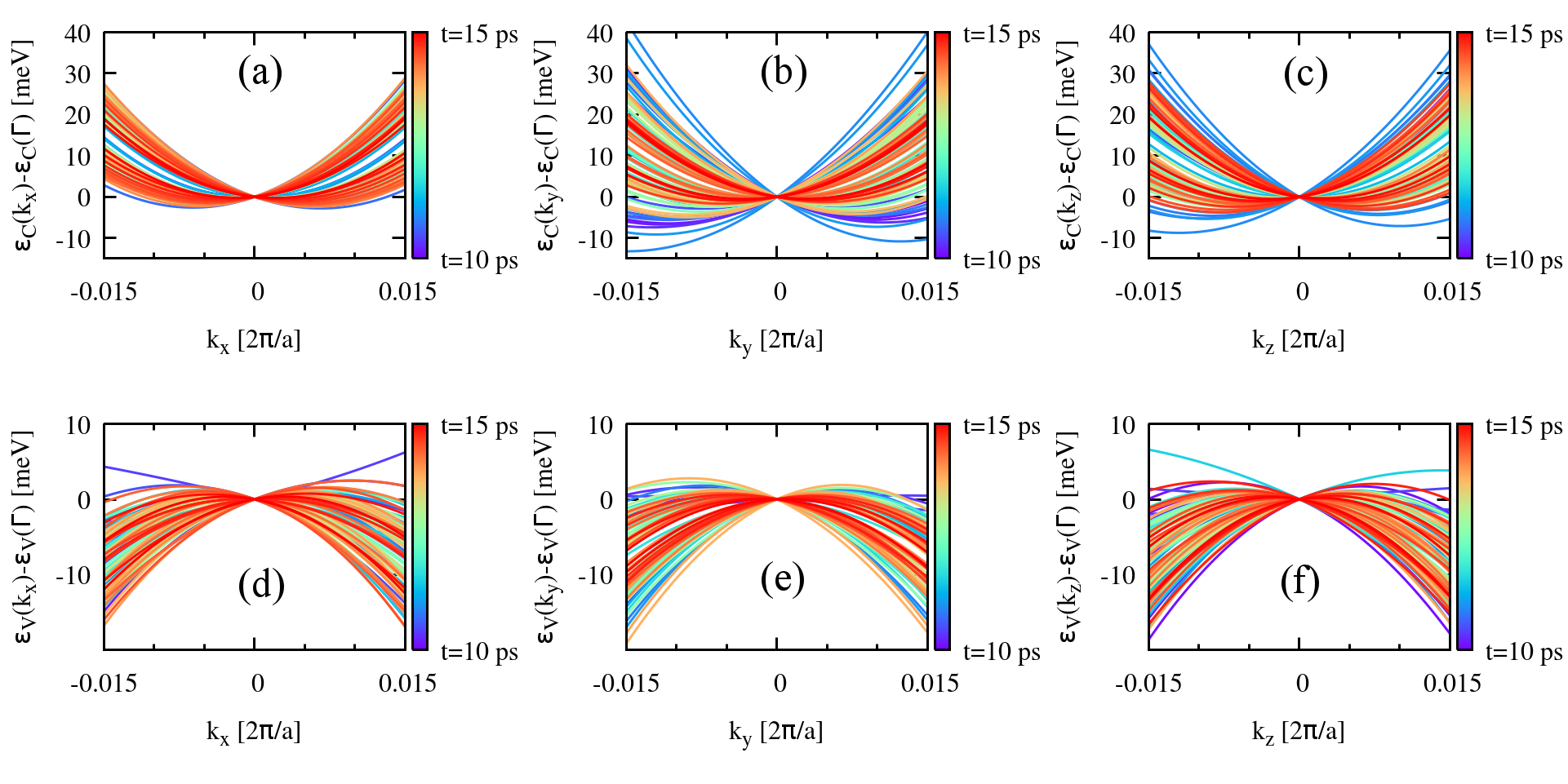}
\caption{Band structure along the three reciprocal space direction, {\bf
k}$_x$, {\bf k}$_y$, {\bf k}$_z$, for the first unoccupied conduction
band (panels a-c) and the last occupied valence one (panels d-e). 
} 
\label{DeltaBands}
\end{center}
\end{figure}

In the main manuscript we showed the band gap around the $\Gamma$ point of the BZ along
the MD simulation. 
The
corresponding highest occupied valence band and lowest unoccupied conduction
band are shown in the six panels (a-f) of figure\ref{DeltaBands}. As for figure 5 of the main manuscript, showing the band gap, 
the bands are referred to the value at $\Gamma$ for the same snapshot, i.e., we show :
$\epsilon_i({\bf k},t)-\epsilon_i(\Gamma,t)$, where $i$ labels the valence band top (VBT) or the conduction band minimum (CBM).

These bands were obtained by assuming a parabolic shape passing through the
three calculated {\bf k} points for each cartesian direction.
The 7 {\bf k}-points ($\Gamma$ belongs to the three directions) are shown in Table~\ref{table_q}

\begin{table}[!h]
\begin{center}
  \begin{tabular}{c | c | c |c |}

    Point number& x &  y & z \\ \hline
    1 & 0.1 & 0 & 0 \\
    2 & 0.05 & 0 & 0 \\
    3 ($\Gamma$) & 0 & 0 & 0 \\
    4 & 0 & 0.05 & 0 \\
    5 & 0 & 0.1 & 0 \\
    6 & 0 & 0 & 0.05 \\
    7 & 0 & 0 & 0.1 \\
  \end{tabular}
\caption{The {\bf k} points of the Brillouin zone used in the 50 band structure calculations in units of $\frac{2\pi}{a^i_{x,y,z}}$ where $a^i_{x,y,z}$ are the lattice parameters for snapshot i in one of the three cartesian direction, x, y or z.}
  \label{table_q}
\end{center}
\end{table}

 It is the assumption of a parabolic shape that induces a slight 
time reversal symmetry breaking ($\epsilon_{\bf k}\neq\epsilon_{-{\bf k}}$). 
It allows to appreciate the small error associated with the calculation of the Rashba parameter $\alpha_R$.

\begin{figure}
\begin{center}
\includegraphics[scale=0.45]{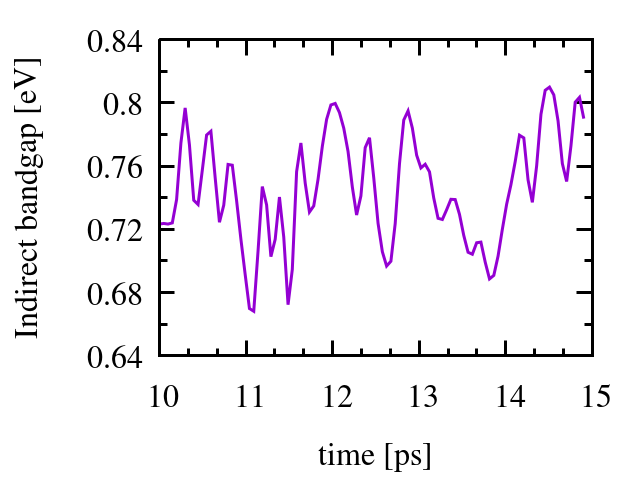}
\caption{Evolution of the band gap in the time interval of 5 ps which was analyzed. } 
\label{GapEvolution}
\end{center}
\end{figure}

The bands are always referred to the value at $\Gamma$, because the time fluctuations of the bands themselves, 
and of the bandgap itself, are much larger than the energy splitting $\Delta E$ entering the definition of the Rashba parameter $\alpha_R$.
We show, as an indication of this, the evolution of the band gap throughout
the 5 ps interval on which we have focused (figure \ref{GapEvolution}).
As can be seen, there is no sign of gap closure during the simulation.

\section{Time correlation functions}
\label{CorrFunct}

\begin{figure}
\begin{center}
   \includegraphics[scale=0.45]{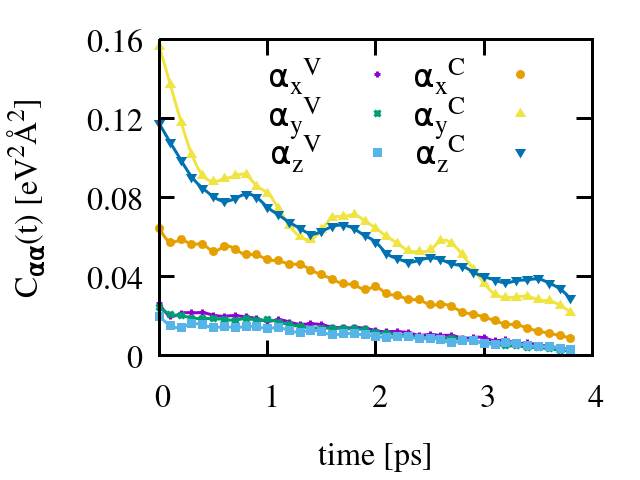}
  \caption{Time autocorrelation function of the Rashba $\alpha_R$ paramenter
    for the three cartesian directions for valence and conduction bands. } 
  \label{corralpha}
  \end{center}
\end{figure}

We further investigated time correlation functions of the Rashba parameter and of spin related functions.
Correlation functions between two functions of time A(t) and B(t) were
calculated, along our 5 ps MD trajectory, as
discrete sums:
$$C_{AB}(j)=\frac{1}{(M+1)}\sum_{i=0}^MA(i)B(i+j)$$
where $i$ and $j$ label the discrete snapshots of the simulation.
We have checked the convergence with the number of points M used for averaging
and found that for M=11 time correlation functions are sufficiently well
converged; nevertheless we are aware that our 50 simulation snapshots represent
a very small statistical sample and allow only qualitative considerations.
First we show in figure \ref{corralpha} the autocorrelation function of the $\alpha_R$ parameter for
the three cartesian directions for valence and conduction bands.

\begin{figure}
\begin{center}
   \includegraphics[scale=0.45]{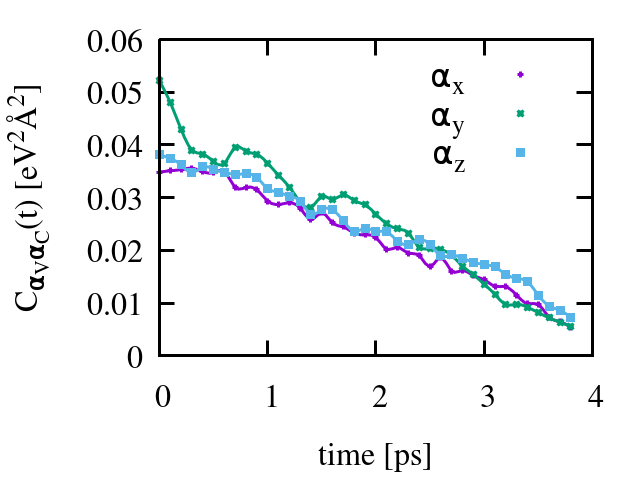}
  \caption{Time cross-correlation function for the valence/conduction Rashba $\alpha_R$ parameters
    for the three cartesian directions. } 
  \label{crossalpha}
  \end{center}
\end{figure}

The fact that the correlation functions show a coherent monotonously
decreasing behavior over a few picoseconds means that the $\alpha_R$ parameter
is still correlated on this time scale. As expected the conduction band shows
stronger Rashba splitting, in particular along ${\bf k}_y$ and ${\bf k}_z$
directions. The largest effect is expected when $\alpha_R$ is large for
both valence and conduction bands. For this reason we calculated the cross
correlation function between the $\alpha_R$ parameters of valence and
conduction bands, C$_{\alpha_R^V\alpha_R^C}$(t), which is shown in figure \ref{crossalpha}.

Here, again, we find a clear decreasing trend, sign of a correlation between
valence and conduction Rashba parameters over a time scale of the same order
as for the autocorrelation functions.

Let us now switch to correlation functions related to the spin texture. We
consider the function shown on figure 2c of the main manuscript; we could call it the
spin-matching-function:
$$S_{match}(t,{\bf k})=\sum_{i=1,3}\langle \sigma_i\rangle_V^{\bf k}\cdot\langle \sigma_i\rangle_C^{\bf k}$$
where the expectation values of the Pauli spin matrices
$\langle\sigma_i\rangle_{V/C}^{\bf k}$ are taken over the spinor
valence/conduction wave function $\psi_{\bf k}^{V/C}({\bf r},t)$, at
point {\bf k} of the BZ and at time $t$.
We have not interpolated the spin texture along the
{\bf k$_{xyz}$} directions, so we consider only the calculated {\bf k} points. 

\begin{figure}
\begin{subfigure}{0.48\textwidth}%
   \includegraphics[scale=0.35]{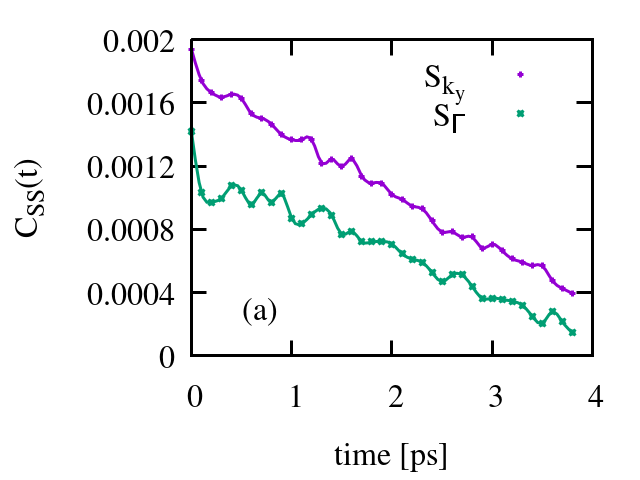}\label{SpinAuto}
\end{subfigure}
\begin{subfigure}{0.48\textwidth}
   \includegraphics[scale=0.35]{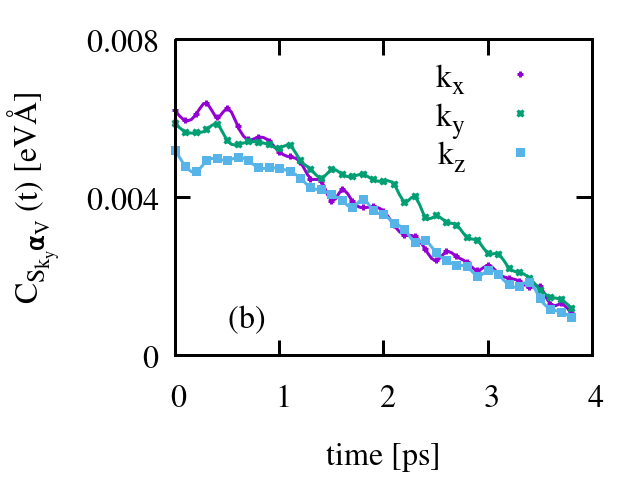}\label{SpinCrossV}
\end{subfigure}
\begin{subfigure}{0.48\textwidth}
   \includegraphics[scale=0.35]{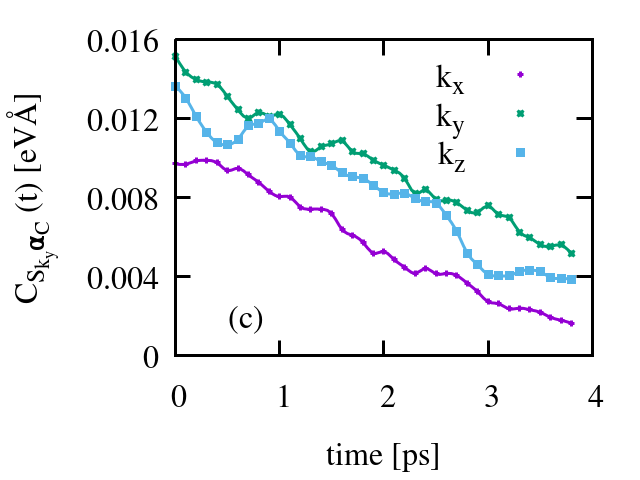}\label{SpinCrossC}
\end{subfigure}
\caption{Panel (a): time autocorrelation function of the spin matching function
  (see text)
  S$_{match}$ at $\Gamma$ and a point along {\bf k}$_y$. Panels (b,c) : cross correlation functions between 
 the spin-matching-function and the  Rashba $\alpha_R$ parameter of
the valence (b) and conduction (c) bands.}
\label{auto-cross-Smatch}
\end{figure}

We have checked the autocorrelation functions for S$_{match}$(t,{\bf k}) at the
$\Gamma$ point and at one {\bf k} point in each of the cartesian directions,
as well as the average of S$_{match}$ over the 7 calculated {\bf k}-points: in
all cases a similar decreasing trend is observed. The same holds for the
cross correlation functions between $\alpha_R$ and S$_{match}$. Examples are
shown in figure \ref{auto-cross-Smatch}.

\begin{figure}
   \includegraphics[scale=0.35]{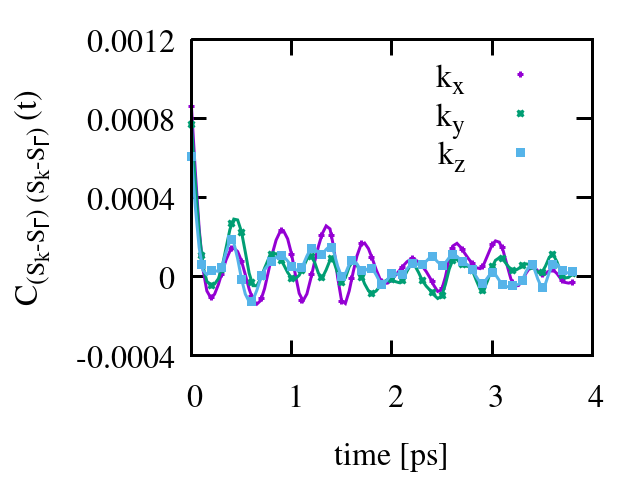}
\caption{Time autocorrelation function of the difference between
  spin-matching-functions at $\Gamma$ and at a finite {\bf k} point along the
  three cartesian directions in the BZ.}
\label{SpinAutoDelta}
\end{figure}

However, we are interested in knowing if the spin matching function, and thus
presumably the recombination rate, is reduced at a finite {\bf k} point with
respect to the zone center. For this we consider $S_{match}(t,{\bf k})-S_{match}(t,\Gamma)$; the corresponding time autocorrelation function is shown in figure
\ref{SpinAutoDelta}, which clearly shows that this difference gets uncorrelated in a much
shorter time than the $\alpha_R$ parameter. 
This suggests that either this reduction is too small to be appreciated,
and/or it is indeed completely uncorrelated with the Rashba effect.
 
